# Bayesian Lifetime Regression with Multi-type Group-shared Latent Heterogeneity

Xuxue Sun[1], Mingyang Li[1]

[1]Department of Industrial and Management Systems Engineering, University of South Florida, USA

**Abstract**

Products manufactured from the same batch or utilized in the same region often exhibit correlated lifetime observations due to the latent heterogeneity caused by the influence of shared but unobserved covariates. The unavailable group-shared covariates involve multiple different types (e.g., discrete, continuous, or mixed-type) and induce different structures of indispensable group-shared latent heterogeneity. Without carefully capturing such latent heterogeneity, the lifetime modeling accuracy will be significantly undermined. In this work, we propose a generic Bayesian lifetime modeling approach by comprehensively investigating the structures of group-shared latent heterogeneity caused by different types of group-shared unobserved covariates. The proposed approach is flexible to characterize multi-type group-shared latent heterogeneity in lifetime data. Besides, it can handle the case of lack of group membership information and address the issue of limited sample size. Bayesian sampling algorithm with data augmentation technique is further developed to jointly quantify the influence of observed covariates and group-shared latent heterogeneity. Further, we conduct comprehensive numerical study to demonstrate the improved performance of proposed modeling approach via comparison with alternative models. We also present empirical study results to investigate the impacts of group number and sample size per group on estimating the group-shared latent heterogeneity and to demonstrate model identifiability of proposed approach for different structures of unobserved group-shared covariates. We also present a real case study to illustrate the effectiveness of proposed approach.

**Keywords:** Group-shared latent heterogeneity, Group-shared unobserved covariates, Lifetime model, Bayesian estimation, Data augmentation

# 1. Introduction

Product reliability analysis is helpful to facilitate life cycle assessment of engineering components and systems. Typically, the lifetime data (e.g., product failure time) in field operation stage provides useful information about product reliability. An accurate predictive model of the product lifetime is of great importance for achieving proactive and effective maintenance policy [1]. The conventional approaches often assume that the lifetime observations are independent and identically distributed. However, this assumption is no longer valid in the application areas of modern industry [2-4]. For example, the failure time observations of products manufactured in same batch are correlated and dependent due to similar manufacturing process conditions (e.g., temperature, humidity, pressure), while they can be dramatically different from the other batches [3]. Such violation against the assumption of independent and uncorrelated samples refers to the intra-group dependencies. The intra-group dependencies quantified by group-specific heterogeneity can be partially explained by observed covariates. Nevertheless, the remaining unexplained part is rather substantial and can only be described by group-shared (e.g., batch-specific, region-specific) unobserved covariates [3]. In this work, we use the term *group-shared latent heterogeneity* to represent the influence of group-shared unobserved covariates on lifetime data. The group-shared unobserved covariates carry important information about product reliability in design and manufacturing stage as well as in field operation stage. Ignoring such information will induce modeling inaccuracy and lead to wrong conclusions about physical nature of products. There is a need to develop a modeling approach with group-shared latent heterogeneity and estimation techniques to account for the influence of group-shared unobserved covariates. A comprehensive understanding of group-shared latent heterogeneity is mandatory for cost effective decisions in both design and production stage as well as in field operation stage. In design and manufacturing stage, successful modeling of group-shared latent heterogeneity can provide quality metrics to product suppliers and deliver useful information to manufacturer about effective design settings and manufacturing variables. This further can help engineers to produce high quality and reliability products. In field operation stage, adequate quantification of group-shared latent heterogeneity can improve the accuracy of product reliability evaluation in field operating environment and further facilitate condition-based maintenance decisions.

Better modeling of group-shared latent heterogeneity requires a basic understanding of the source of the influencing group-shared unobserved covariates. The unobserved covariates shared within group mainly originate from the following sources in real world. Some of the influencing factors of product reliability are known and explanatory (e.g., material quality indicators, design settings, manufacturing process variables), and they are shared in same group (e.g., same batch in production stage, same region in operation stage). However, they are not available due to the privacy and ownership issue of company warranty data [4]. In addition, the limited sensing techniques and measurement capability as well as high costs make the covariates unavailable, such as operating environment conditions in field operation stage [5]. Moreover, the unobserved covariates can also be unknown/unexplained factors due to the limited knowledge available for new material products [6]. In this work, the data quality issue attributed to the unavailability of covariates information is the main focus and will be addressed via comprehensive investigation.

In addition to the source of group-shared unobserved covariates, it is essential to comprehensively explore the possible structures of such covariates information and further develop an effective modeling approach for characterizing the group-shared latent heterogeneity. There are three main types of covariates information, including qualitative type, quantitative type, and mixed-type [7, 8]. The qualitative covariates can be nominal, such as different types of material descriptors and different design configurations [4, 7], or ordinal, such as different levels of material quality and different use conditions [8, 9]. Such qualitative type of covariates information can be represented by discrete random variables. Moreover, the covariates information can be quantitative and represented by continuous random variables. The quantitative covariates can be the continuous variables for process conditions (e.g., pressure, ambient temperature) in manufacturing stage [7, 10, 11], or the continuous variables for environmental operating conditions (e.g., loadings, temperature, humidity) in field operation stage [8, 12]. Further, the covariates information can also become mixed-type (e.g., both qualitative and quantitative) and represented by a combination of continuous and discrete random variables [13, 14]. When such complex and multi-type covariates are shared within same group (e.g., same batch during manufacturing, same region in field operation) and unavailable due to data quality issue, these covariates can be specified as group-shared unobserved covariates and their structures involve multiple different types. The complex structures of group-shared unobserved covariates raise the challenges to the quantification of multi-type group-shared latent heterogeneity in lifetime data analysis.

In the existing literature of product lifetime modeling, most of them only considered the observed heterogeneity [15, 16]. Many conventional approaches were developed under the assumption that the lifetime observations within same group are independent. They often used distribution-based methods to characterize the heterogeneity of product lifetime as a whole, or utilized regression-based models to quantify the influence of observed covariates alone. Although the influence of unobserved covariates was considered in some recent studies, these existing methods mainly focused on capturing the intra-individual dependency of product lifetime in a repairable system. A continuous random variable was often introduced to explain part of individual heterogeneity and to represent unit-to-unit variation. These lifetime models with continuous latent variable include frailty model [17] and its multivariate variation [18] in survival analysis, and models of failure time data which consider the influence of unknown use conditions in field operation stage [19]. Among the existing approaches based on frailty model, different types of continuous distribution can be specified for the frailty term, such as Gamma distribution [20-22], and generalized inverse Gaussian distribution [23]. Other analytical models are also developed to investigate the continuous type of individual heterogeneity of product lifetime, such as continuous stochastic process with random coefficients [24]. Nevertheless, few studies in product reliability analysis are found to address the intra-group dependency and to capture the influence of group-shared unobserved covariates. Further, limited studies account for the discrete type and mixed-type of latent heterogeneity, especially at group level. There is a gap between the existing approaches and the goal of improved accuracy of lifetime modeling in presence of different structures of group-shared unobserved covariates. It becomes mandatory to resolve the within-group dependency of lifetime data [25] due to the unavailability of group-shared covariates and to address the data quality issues via statistical modeling approach. Moreover, the conventional estimation methods, such as maximum likelihood estimation [26], cannot be applied to estimate the group-shared latent heterogeneity because the latent variables will be integrated out via the marginal approach. Besides, the lifetime observations could become independent given that they are conditioned on the group-shared latent heterogeneity, which also increases the difficulty to capture the intra-group dependency of lifetime data.

To fill the above research gaps, we propose a generic modeling approach to characterize the lifetime data (e.g., product failure time) with both observed heterogeneity and multi-type group-shared latent heterogeneity. Specifically, we comprehensively investigate different structures of group-shared unobserved covariates and different types of their influences, namely, discrete group-shared latent heterogeneity (GSLH-D), continuous group-shared latent heterogeneity (GSLH-C) and mixed-type (e.g., both continuous and discrete types) group-shared latent heterogeneity (GSLH-M). According to the principle of missing information [27], a flexible dependence structure is introduced to capture the latent heterogeneity shared within group. Besides, the proposed model is flexible to analyze failure time data with different specifications of accelerated failure time (AFT) model (e.g., Log-normal, Weibull). Further, estimation algorithm under Bayesian framework is developed to jointly quantify both the observed heterogeneity and multi-type group-shared latent heterogeneity. The developed estimation procedure is able to address the intra-group dependency while membership information and considerable sample size in each group are not required. Moreover, we conduct comprehensive comparison study to demonstrate the effectiveness and model identifiability of the proposed approach.

The rest of this paper is organized as follows. In Section. 2, we present the model details of the proposed approach with multi-type group-shared latent heterogeneity. In Section. 3, model estimation details are elaborated to account for different structures of group-shared unobserved covariates. In Section. 4, we first conduct a numerical study to compare the prediction performance of proposed approach with other alternative approaches. We also investigate the quantification of different types of group-shared latent heterogeneity via empirical analysis. Then, we present a real study to demonstrate the effectiveness of proposed approach and illustrate its ability of identifying the structure of group-shared unobserved covariates. In Section. 5, we draw the conclusive remarks of this work.

## 2. Model Formulation

Consider a heterogeneous population of non-repairable units in $n$ groups (e.g., a population of product items manufactured in $n$ batches), there are $m_i$ lifetime observations (e.g. product failure time) within each group $i, \forall i = 1 \dots n$. We can employ AFT model to analyze the lifetime data due to its modeling flexibility and sound model interpretability. The general form of AFT model can be expressed as

$$log(t_{ij}) = \beta_0 + \boldsymbol{\beta}^{\mathrm{T}} \boldsymbol{x}_{ij} + \epsilon_{ij}, \quad i = 1, \dots, n, \ j = 1, \dots, m_i \quad (1)$$

where $\boldsymbol{x}_{ij}$ is a vector of covariates of $j$th unit in $i$th group and $\boldsymbol{\beta}$ is a vector of corresponding coefficients at logarithmic scale. $\beta_0$ is the average lifetime in absence of the covariates at logarithmic scale. $\varepsilon_{ij}$ represents the measurement error of $j$th unit in $i$th group, which is assumed to be independent with zero mean and bounded variance. With different assumptions on $\varepsilon_{ij}$, the AFT model can be specified as different regression models, such as Log-normal regression model and Weibull regression model. These two specifications of AFT model are considered in this work.

Further, we manifest the components of the covariates $\boldsymbol{x}_{ij}$ to characterize the intra-group dependency. The covariates $\boldsymbol{x}_{ij}$ embrace the observed covariates $\widetilde{\boldsymbol{x}}_{ij}$, such as stress factor and operating temperature, and the unobserved covariates, such as manufacturing variables which are not available from warranty data due to data privacy issue. The unobserved covariates contribute to the latent heterogeneity of product lifetime. In this work, we focus on the lifetime modeling of non-repairable products and thus the individual latent heterogeneity due to correlated and repeated measurements becomes negligible. Thus, the latent heterogeneity only consists of the group-shared latent heterogeneity induced by the unobserved covariates $Z_i$ shared within each group $i$ (e.g., unavailable material quality variables of suppliers in same region, unavailable manufacturing process variables in same batch). The proposed model can then be expressed as

$$\log(t_{ij}) = \beta_0 + \boldsymbol{\beta}^\mathrm{T} \tilde{\boldsymbol{x}}_{ij} + \boldsymbol{\alpha}^\mathrm{T} \boldsymbol{Z}_i + \epsilon_{ij}, \quad i = 1, \ldots, n \; j = 1, \ldots, m_i \qquad (2)$$

where $\boldsymbol{Z}_i$ represents group-shared unobserved covariates for group $i$ and $\boldsymbol{\alpha}$ is a vector of corresponding coefficients at logarithmic scale. $\boldsymbol{\beta}^\mathrm{T} \tilde{\boldsymbol{x}}_{ij}$ captures the influence of observed covariates on the lifetime of unit $j$ in group $i$ at logarithmic scale. Further, we denote $W_i = \boldsymbol{\alpha}^\mathrm{T} \boldsymbol{Z}_i$ to quantify the group-shared latent heterogeneity. Different structures can be specified for $\boldsymbol{Z}_i$ and consequently $W_i$ becomes multi-type random quantity.

(i) All $\boldsymbol{Z}_i$'s are qualitative random factors with $K$ discrete values, such as the unavailable material quality descriptors with K-levels from different suppliers in same region. $W_i = \boldsymbol{\alpha}^\mathrm{T} \boldsymbol{Z}_i$ can then be specified as categorical random variable, i.e., $W_i \sim Categ(K, \boldsymbol{p})$ where $Categ(K, \boldsymbol{p})$ refers to categorical distribution with $K$ discrete values, i.e., $d_k, \forall k = 1 \ldots K$, and a vector of probabilities $\boldsymbol{p} = [p_1, \ldots, p_K]^\mathrm{T}$ such that $\sum_{k=1}^K p_k = 1$. Such discrete type of group-shared latent heterogeneity is annotated as Type I (GSLH-D).

(ii) All $\boldsymbol{Z}_i$'s are quantitative random factors, such as the unavailable ambient temperature of product items in same batch during manufacturing process. $W_i = \boldsymbol{\alpha}^\mathrm{T} \boldsymbol{Z}_i$ then becomes continuous random variable, i.e., $W_i \sim G(\cdot)$ where $G(\cdot)$ represents continuous density. Such continuous type of group-shared latent heterogeneity is annotated as Type II (GSLH-C).

(iii) $\boldsymbol{Z}_i$ is a combination of both qualitative and quantitative factors, such as the unavailable region-shared categorical hotness degree and other continuous environmental conditions in field operation. $W_i = \boldsymbol{\alpha}^\mathrm{T} \boldsymbol{Z}_i$ becomes mixed-type (both continuous and discrete) random variable, i.e., $W_i \sim G_k(\cdot)$ with probability $q_k, \forall k = 1 \ldots K$ such that $\sum_{k=1}^K q_k = 1$. $G_k(\cdot)'s$ represent K different continuous densities. Such mixed-type group-shared latent heterogeneity is annotated as Type III (GSLH-M).

Attributed to the multi-type group-shared latent heterogeneity $W_i$, the proposed modeling approach is generic and flexible to capture the influence of group-shared unobserved covariates with different structures. In addition, the proposed lifetime model is free of any assumptions on the concrete form of failure hazard. The conventional frailty model (with continuous latent variable) [17,18] and mixture model (with discrete latent variable) [28,29] can be perceived as special cases of proposed model. We will elaborate the details of estimation procedure as follows.

## 3. Model Estimation

Suppose $m_i$ lifetime observations within group $i, \forall i = 1 \ldots n$ (e.g., product failure time data in same batch), are collected together with the observed covariates $\tilde{\boldsymbol{x}}_{ij}, \forall i = 1 \ldots n, j = 1 \ldots m_i$. We introduce a binary indicator $\delta_{ij}$ to account for the right-censored $j^{\text{th}}$ lifetime observation in $i^{\text{th}}$ group. When $\delta_{ij}$ takes value 1, $t_{ij}$ represents the elapsed time before the critical event of product failure occurs within the time period of data collection. Otherwise, $t_{ij}$ represents the whole period of data collection with $\delta_{ij} = 0$. The available data can then be represented as $\boldsymbol{D} = \{t_{ij}, \delta_{ij}, \widetilde{\boldsymbol{x}}_{ij}, \forall i = 1 \dots n, j = 1 \dots m_i\}$. We denote a set of unknown model parameters as $\boldsymbol{\Theta}$. The marginal likelihood function $L(\boldsymbol{\Theta} \mid \boldsymbol{D})$ can be expressed as

$$L(\Theta \mid D) = \prod_{i=1}^{n} \int_0^{\infty} \prod_{j=1}^{m_i} [f(t_{ij} \mid \Theta, W_i)]^{\delta_{ij}} \cdot [R(t_{ij} \mid \Theta, W_i)]^{1-\delta_{ij}} \cdot f_w(W_i) dW_i \quad (3)$$

where $f(t_{ij} \mid \boldsymbol{\Theta}, W_i)$ is the density function for lifetime of product unit $j$ in group $i$ and $R(t_{ij} \mid \boldsymbol{\Theta}, W_i) = 1 - \int_0^{t_{ij}} f(s \mid \boldsymbol{\Theta}, W_i) ds$ is the reliability function. $f_w(\cdot)$ is the density function for the group-shared latent heterogeneity. In the conventional non-Bayesian estimation methods, such as maximization likelihood estimation, $W_i$'s will be integrated out and cannot be estimated. Nevertheless, $W_i$'s carry all important information for the quantification of group-shared latent heterogeneity. To overcome the limitation of conventional methods, Bayesian estimation framework is adopted to develop the estimation algorithm due to its estimation power and flexibility. Both of the unknown parameters $\boldsymbol{\Theta}$ and all $W_i$'s can then be jointly estimated, and exact inference of $\boldsymbol{\Theta}$ and $W_i$'s can be obtained. We denote the joint prior density for unknown parameters as $\pi(\boldsymbol{\Theta})$, which indicates the available prior knowledge of all unknown parameters. Further, we denote the hyper-parameters for $W_i$'s as $\boldsymbol{\Phi}$ and let $\pi(W_i \mid \boldsymbol{\Phi})$ be the prior density for $W_i, \forall i = 1, \dots, n$. The joint posterior can then be derived as

$$\pi(\Theta, \{W_i\}_{i=1}^n, \Phi \mid D) \propto L(\Theta, \{W_i\}_{i=1}^n \mid D) \cdot \pi(\Theta) \cdot \prod_{i=1}^{n} \pi(W_i \mid \Phi) \pi(\Phi)$$

$$= \prod_{i=1}^{n} \prod_{j=1}^{m_i} L_{ij}(\Theta, W_i \mid D_{ij}) \cdot \pi(\Theta) \cdot \prod_{i=1}^{n} \pi(W_i \mid \Phi) \pi(\Phi) \quad (4)$$

where $L(\boldsymbol{\Theta}, \{W_i\}_{i=1}^n \mid \boldsymbol{D})$ is the joint likelihood. $L_{ij}(\boldsymbol{\Theta}, W_i \mid D_{ij})$ can be further expressed as $L_{ij}(\boldsymbol{\Theta}, W_i \mid D_{ij}) = f(t_{ij} \mid \boldsymbol{\Theta}, W_i)^{\delta_{ij}} \cdot R(t_{ij} \mid \boldsymbol{\Theta}, W_i)^{1-\delta_{ij}}$ where $D_{ij}$ is the available data for $j^{\text{th}}$ product unit in $i^{\text{th}}$ group.

In this study, we focus on characterizing the intra-group dependency and thus we assume the group-shared latent heterogeneity is not correlated among different groups. In the following subsections, we will elaborate the details of obtaining exact inferences of both unknown parameters $\boldsymbol{\Theta}$ and $W_i$'s as well as hyper-parameters $\boldsymbol{\Phi}$ for each type of group-shared latent heterogeneity.

### 3.1. Type I Group-shared Latent Heterogeneity: GSLH-D

We first investigate the discrete structure of group-shared unobserved covariates. The group-shared latent heterogeneity can be represented by discrete random quantity. Specifically, $W_i$ follows categorical distribution with $K$ possible discrete values $d_k$ and a vector of probability values $\boldsymbol{p} = [p_1, \dots, p_K]^{\text{T}}$, i.e., $P(W_i = d_k) = p_k, \forall k = 1, \dots, K$ such that $\sum_{k=1}^{K} p_k = 1$. The density function for $W_i$ can be written as $f_w(W_i \mid \boldsymbol{p}) = \prod_{k=1}^{K} p_k^{I(W_i = d_k)}$ where $I(\cdot)$ is indicator function. The marginal density of lifetime can then be expressed as $f(t_{ij} \mid \boldsymbol{\Theta}) = \sum_{k=1}^{K} p_k f_k(t_{ij} \mid \boldsymbol{\Theta}, W_i = d_k)$ and the conditional density becomes $f(t_{ij} \mid \boldsymbol{\Theta}, W_i) = \prod_{k=1}^{K} f_k(t_{ij} \mid \boldsymbol{\Theta}, W_i = d_k)^{I(W_i = d_k)}$. The joint posterior density can then be specified as

$$\pi(\Theta, \{W_i\}_{i=1}^n, p \mid D) \propto \prod_{i=1}^n \prod_{j=1}^{m_i} [\prod_{k=1}^K (f_k(t_{ij} \mid \Theta, W_i = d_k))^{I(W_i=d_k)}]^{\delta_{ij}}$$

$$\cdot [1 - \int_0^{t_{ij}} \prod_{k=1}^K f_k(s \mid \Theta, W_i = d_k)^{I(W_i=d_k)} ds]^{1-\delta_{ij}} \cdot \pi(\Theta) \cdot \prod_{i=1}^n f_w(W_i \mid p)\pi(p)$$

$$\propto \prod_{j=1}^{m_i} \prod_{k=1}^K \prod_{i \in s_k} [(f_k(t_{ij} \mid \Theta, W_i = d_k))^{\delta_{ij}} [R_k(t_{ij} \mid \Theta, W_i = d_k)]^{1-\delta_{ij}}$$

$$\cdot \pi(\Theta) \cdot \prod_{k=1}^K p_k^{|s_k|} \pi(p) \tag{5}$$

where $s_k = \{i: W_i = d_k\}, \forall k = 1, \ldots, K$ is the index set of subpopulation k. The size of index set is obtained as $|s_k| = \sum_{i=1}^n I(W_i = d_k)$ such that $\sum_{k=1}^K |s_k| = n$ where $|\cdot|$ is the size operator. $R_k(t_{ij} \mid \Theta, W_i = d_k) = 1 - \int_0^{t_{ij}} f_k(s \mid \Theta, W_i = d_k) ds$ is reliability function of lifetime for the units in all groups which belong to same subpopulation $k$. Further, we can assign Dirichlet conjugate prior to hyper-parameters $p$, i.e., $p \sim Dirichlet(v)$ where $v = [v_1, \ldots, v_K]^T$ is a vector of parameters for Dirichlet distribution. The conditional posterior density for hyper-parameters $p$ can then be obtained as

$$\pi(p \mid \Theta, D, \{W_i\}_{i=1}^n) \propto \prod_{i=1}^n f_w(W_i \mid p)\pi(p) \propto \prod_{k=1}^K p_k^{|s_k|+v_k-1} \tag{6}$$

Based on above equation, the conditional posterior of $p$ is Dirichlet distribution with parameters $v + b$ where $b == [|s_1|, \ldots, |s_K|]^T$. The conditional posterior density for $W_i$'s which belong to same subpopulation $k$ can be derived as

$$\pi(\{W_i\}_{\forall i: W_i = d_k} \mid \Theta, D, p) \propto \prod_{i \in s_k} \left[\prod_{j=1}^{m_i} L_{ij}(\Theta, W_i = d_k \mid D_{ij})\right] \cdot \pi(W_i = d_k \mid p) \tag{7}$$

For any of $W_i$ which takes value $d_k$, it is conditionally independent of $W_j, \forall j \notin s_k$. With such less correlated structure and reduced computational complexity, the samples of $W_i$'s can be generated efficiently.

Different regression models can be further specified for Equation. 2 based on different assumptions on the distribution of random error $\varepsilon_{ij}$. In this study, standard normal distribution and extreme value distribution are considered. The proposed model can be transformed to Log-normal regression model and Weibull regression model respectively. With different specifications, the full conditional posterior of unknown parameters can be obtained as well. For example, consider a Weibull regression model with rate parameter $\lambda$, the full conditional posterior density of $\lambda$ can be derived as

$$\pi(\lambda \mid \Theta^{-\lambda}, D, \{W_i\}_{i=1}^n) \propto \prod_{k=1}^K \left[\prod_{i \in s_k} \prod_{j=1}^{m_i} L_{ij}(\Theta, W_i = d_k \mid D_{ij})\right] \cdot \pi(\lambda) \tag{8}$$

Since $\lambda > 0$, gamma prior can be assigned. As no conjugate prior is readily available, we can adopt Metropolis-Hasting algorithm [30] to generate the posterior samples of $\lambda$.

Moreover, with reparameterization technique, the proposed model with Type I group-shared latent heterogeneity can be reduced to the mixture regression model [28,29]. The following proposition clarifies the relationship between the mixture regression model and the proposed model.

**Proposition 1.** Under different specification for $\varepsilon_{ij}$, the proposed model with discrete type of group-shared latent heterogeneity, as shown in Section. 2, can be reduced to Weibull mixture regression models and Log-normal mixture regression models with specific unknown parameters $\Theta_k$ for each subpopulation $k$, $\forall k = 1, \ldots, K$, i.e., $f(t \mid \Theta) = \sum_{k=1}^{K} p_k f_k(t \mid \Theta_k)$ where $f_k(\cdot)$ is density function (e.g., Weibull density and Log-normal density). Specifically,

1. suppose $\varepsilon_{ij}$ follows normal distribution, i.e., $\varepsilon_{ij} = \sigma \eta_{ij}$ where $\eta_{ij}$ follows standard normal distribution and $\sigma > 0$. Further, we manifest the subpopulation-specific unknown parameters as $\Theta_k = \{\mu_k, \sigma_k^2\}$ where $\mu_k$ and $\sigma_k^2$ are mean and variance at logarithmic scale respectively for product lifetime which belongs to subpopulation $k$. Then, the Log-normal mixture regression model, i.e., $f(t) = \sum_{k=1}^{K} p_k LN(\mu_k = \beta_0 + \boldsymbol{\beta}^T \tilde{x}_{ij} + d_k, \sigma_k^2 = \sigma^2)$, becomes equivalent to the proposed model with Type I group-shared latent heterogeneity.

2. suppose $\varepsilon_{ij}$ follows extreme distribution, i.e., $\varepsilon_{ij} = \sigma \eta_{ij}$ where $\eta_{ij}$ follows standard gumbel distribution and $\sigma > 0$. Further, we manifest the subpopulation-specific unknown parameters as $\Theta_k = \{\lambda_k, \rho_k\}$ where $\lambda_k$ is rate parameter and $\rho_k$ is shape parameter for product lifetime which belongs to subpopulation $k$. Then, the Weibull mixture regression model, i.e., $f(t) = \sum_{k=1}^{K} p_k Weib(\lambda_k = \exp[-\frac{1}{\sigma}(\beta_0 + \boldsymbol{\beta}^T \tilde{x}_{ij} + d_k)], \rho_k = \frac{1}{\sigma})$, becomes equivalent to the proposed model with Type I group-shared latent heterogeneity.

The proposition implies that the baseline hazard becomes subpopulation-specific instead of being shared among different subpopulations. The Type I group-shared latent heterogeneity can be quantified by the subpopulation-specific parameters $\Theta_k$. The proof details of the proposition can be found in Appendix A.

*3.2. Type II Group-shared Latent Heterogeneity: GSLH-C*

Further, we investigate the continuous structure of group-shared unobserved covariates. The group-shared latent heterogeneity can be captured by continuous random quantity, i.e., $W_i \sim G(\cdot)$ where $G(\cdot)$ represents arbitrary continuous density function with hyper-parameters $\Phi$. When the group-shared latent heterogeneity is Type II, the proposed model in Equation.2 is reduced to the frailty model [17] and the frailty term can be quantified by $\exp(W_i)$. The joint posterior density then becomes

$$\pi(\Theta, \{W_i\}_{i=1}^{n}, \Phi \mid D) \propto \prod_{i=1}^{n} \prod_{j=1}^{m_i} L_{ij}(\Theta, W_i \mid D_{ij}) \cdot \pi(\Theta) \cdot \prod_{i=1}^{n} G(W_i \mid \Phi) \pi(\Phi) \quad (9)$$

The estimates of unknown parameters $\Theta$ and $W_i$'s as well as hyper-parameters $\Phi$ can be derived directly. The full conditional posteriors of $W_i$ and $\Phi$ can be expressed as

$$\pi(W_i \mid \Theta, \Phi, D) \propto \prod_{j=1}^{m_i} L_{ij}(\Theta, W_i \mid D_{ij}) \cdot G(W_i \mid \Phi)$$

$$\pi(\Phi \mid \{W_i\}_{i=1}^{n}, \Theta, D) \propto \prod_{i=1}^{n} G(W_i \mid \Phi) \cdot [\pi(\Phi)]^n \quad (10)$$

## 3.3. Type III Group-shared Latent Heterogeneity: GSLH-M

The above GSLH-D model and GSLH-C model can be perceived as special cases of the GSLH-M model, with only discrete component and only continuous component respectively. In the GSLH-M model, the group-shared unobserved covariates are combination of both discrete and continuous quantities. The group-shared latent heterogeneity can be represented by mixed-type random variable, i.e., $W_i \sim G_k(\cdot)$ with probability $q_k$ such that $\sum_{k=1}^{K} q_k = 1$ and each $G_k(\cdot)$ represents continuous density with parameters $\boldsymbol{\phi}_k$ for all $W_i$'s which belong to subpopulation $k, \forall k = 1 \ldots K$. When the whole population consists of only one subpopulation, i.e., $K = 1$, the GSLH-M model is reduced to the GSLH-C model. When the randomness of $W_i$ in each subpopulation diminishes and $W_i$'s become constant values within each subpopulation, the GSLH-M model is reduced to the GSLH-D model. In the GSLH-M model, the joint posterior can be expressed as

$$\pi(\Theta, \{W_i\}_{i=1}^{n}, q, \{\phi_k\}_{k=1}^{K} \mid D) \propto \prod_{i=1}^{n} \prod_{j=1}^{m_i} L_{ij}(\Theta, W_i \mid D_{ij})$$

$$\cdot \pi(\Theta) \cdot \prod_{i=1}^{n} \left( \sum_{k=1}^{K} q_k G_k(W_i \mid \phi_k) \right) \pi(q) \prod_{k=1}^{K} \pi(\phi_k) \quad (11)$$

where $\boldsymbol{q} = [q_1, \ldots, q_K]^T$. Based on the above form of joint posterior density, it becomes mathematically intractable to generate samples for $W_i$'s since the priors for all $W_i$'s involve $K^n$ additive forms. Data augmentation techniques [31] are introduced to address the practical issues of analytical complexity and to improve computational efficiency. We introduce an augmented variable $\xi_i$ to indicate the subpopulation membership of each $W_i, \forall i = 1, \ldots, n$. $\xi_i$ takes value $k$ if group $i$ belongs to subpopulation $k$, i.e., $W_i \mid \xi_i = k \sim G_k(\cdot)$. The density function for $\xi_i$ is written as $f(\xi_i \mid q) = \prod_{k=1}^{K} q_k^{I(\xi_i=k)}$ where $I(\cdot)$ is indicator function. The conditional density of Wi is then written as $f(W_i \mid \xi_i, \{\phi_k\}_{k=1}^{K}) = \prod_{k=1}^{K} G_k((W_i \mid \boldsymbol{\phi}_k)^{I(\xi_i=k)}$. Further, we denote index set of subpopulation $k$ as $I_k = \{i: \xi_i = k\}, \forall k = 1, \ldots, K$ with $|I_k| = \sum_{i=1}^{n} I(\xi_i = k)$ such that $\sum_{k=1}^{K} |I_k| = n$ where $|\cdot|$ is the size operator. The joint posterior can then be derived as

$$\pi(\Theta, \{W_i\}_{i=1}^{n}, q, \{\phi_k\}_{k=1}^{K}, \{\xi_i\}_{i=1}^{n} \mid D,)$$

$$\propto \prod_{i=1}^{n} \prod_{j=1}^{m_i} L_{ij}(\Theta, W_i \mid D_{ij}) \cdot \pi(\Theta) \cdot \prod_{i=1}^{n} f(W_i \mid \xi_i, \{\phi_k\}_{k=1}^{K}) f(\xi_i \mid q) \pi(q) \prod_{k=1}^{K} \pi(\phi_k)$$

$$\propto \prod_{i=1}^{n} \prod_{j=1}^{m_i} L_{ij}(\Theta, W_i \mid D_{ij}) \cdot \pi(\Theta) \cdot \prod_{k=1}^{K} \prod_{i \in I_k} G_k(W_i \mid \phi_k) \cdot \prod_{k=1}^{K} q_k^{|I_k|} \pi(q) \pi(\phi_k) \quad (12)$$

Dirichlet conjugate prior can be assigned for $\boldsymbol{q}$, i.e., $\boldsymbol{q} \sim Dirichlet(\boldsymbol{\zeta})$ where $\boldsymbol{\zeta} = [\zeta_1, \ldots, \zeta_K]^T$ is a vector of parameters for Dirichlet distribution. The full conditional posterior of $\boldsymbol{q}$ can then be obtained as

$$\pi(q \mid \Theta, D, \{\xi_i\}_{i=1}^{n}, \{W_i\}_{i=1}^{n}, \{\phi_k\}_{k=1}^{K}) \propto \prod_{k=1}^{K} q_k^{\zeta_k + |I_k| - 1} \quad (13)$$

The conditional posterior of $\boldsymbol{q}$ has a Dirichlet density with parameter $\boldsymbol{\zeta} + \boldsymbol{c}$ where $\boldsymbol{c} = [|I_1|, \ldots, |I_K|]^T$. The conditional posteriors for $\xi_i$ and $W_i$ can then be derived as

$$\pi(\xi_i \mid W_i, D, q, \{\phi_k\}_{k=1}^{K}) \propto f(W_i \mid \xi_i, \{\phi_k\}_{k=1}^{K}) f(\xi_i \mid q) \propto \prod_{k=1}^{K} [q_k G_k(W_i \mid \phi_k)]^{I(\xi_i=k)}$$

$$\pi(W_i \mid \xi_i = k, \Theta, D, q, \phi_k) \propto \prod_{j=1}^{m_t} L_{ij}(\Theta, W_i \mid D_{ij}) \cdot G_k(W_i \mid \phi_k) \qquad (14)$$

Based on above derivations, we develop a generalized estimation algorithm for the proposed approach, as shown in Algorithm 1. $\tau_{\max}$ is the maximum number of iterations. Without any available

---

**Algorithm 1** Sampling algorithm for the proposed modeling framework

**Initialization:** $\Theta^{(0)}, \{W_i^{(0)}\}_{i=1}^{n}$
  **if** GSLH-D model **then**
    $p^{(0)} \sim \text{Dirichlet}(\nu)$
  **end if**
  **if** GSLH-C model **then** Initialize $\Phi^{(0)}$
  **end if**
  **if** GSLH-M model **then**
    $q^{(0)} \sim \text{Dirichlet}(\zeta)$ and initialize $\{\phi_k^{(0)}\}_{k=1}^{K}$
  **end if**
  **procedure** DRAWSAMPLES
    **for** $\tau \leftarrow 1, ..., \tau_{\max}$ **do**
      **if** GSLH-D model **then**
        Draw $W_i^{(\tau)}$ from $\pi(W_i \mid \Theta^{(\tau-1)}, D, p^{(\tau-1)}), \forall i = 1, ..., n$
        Partition data with $s_k^{(\tau)} = \{i : W_i^{(\tau)} = d_k\}, \forall k = 1, ..., K$
        Draw $p^{(\tau)}$ from $\text{Dirichlet}(\nu + b^{(\tau)})$
      **end if**
      **if** GSLH-C model **then**
        Draw $W_i^{(\tau)}$ from $\pi(W_i \mid \Theta^{(\tau-1)}, D, \Phi^{(\tau-1)}), \forall i = 1, ..., n$
        Draw $\Phi^{(\tau)}$ from $\pi(\Phi \mid \{W_i^{(\tau)}\}_{i=1}^{n}, \Theta^{(\tau-1)}, D)$
      **end if**
      **if** GSLH-M model **then**
        Draw $\xi_i^{(\tau)}$ from $\pi(\xi_i \mid W_i^{(\tau-1)}, D, q^{(\tau-1)}), \forall i = 1, ..., n$
        Partition data with $I_k^{(\tau)} = \{i : \xi_i^{(\tau)} = k\}, \forall k = 1, ..., K$
        Draw $W_i^{(\tau)}$ from $\pi(W_i \mid \xi_i^{(\tau)}, \Theta^{(\tau-1)}, D, q^{(\tau-1)}), \forall i = 1, ..., n$
        Draw $q^{(\tau)}$ from $\text{Dirichlet}(\zeta + c^{(\tau)})$
        Draw $\phi_k^{(\tau)}$ from $\pi(\phi_k \mid \Theta^{(\tau-1)}, D, q^{(\tau)}, \{\xi_i^{(\tau)}\}_{i=1}^{n}, \{W_i^{(\tau)}\}_{i=1}^{n}), \forall k = 1, ..., K$
      **end if**
      Draw $\Theta^{\tau}$ from $\pi(\Theta^{\tau} \mid \{W_i^{(\tau)}\}_{i=1}^{n}, D)$
    **end for**
  **end procedure**

---

prior knowledge of $p$ (or $q$), the prior can be elicited from non-informative Dirichlet conjugate prior, e.g., Jefferys prior with $\nu = 0.5$ (or $\zeta = 0.5$) [28], which indicates the structure proportions of subpopulations. For the sampling of unknown parameters $\Theta$ and random quantities $W_i$'s, the posterior samples can be drawn conveniently if conjugate prior is available. Otherwise, Metropolis-Hasting algorithm [30] can be applied to generate the posterior samples.

## 4. Case Study

### 4.1. Numerical Case Study

To illustrate the effectiveness of proposed modeling approach, we first provide simulation study with ground-truth settings to comprehensively investigate different types of group-shared latent heterogeneity and to compare the performance of proposed approach with alternative approaches. In this work, we consider two subpopulations (K=2) for simplicity in the case of mixed-type group shared latent heterogeneity. It can be easily extended to multiple subpopulations via using Dirichlet distribution instead of Beta distribution. In each type of group-shared latent heterogeneity, the continuous component is generated randomly based on normal distribution while the discrete component is generated randomly with a total proportion of 0.35 for subpopulation 1. We consider two different model specifications without losing generality, including Log-normal and Weibull models. Figure.1 shows different structures of group-shared unobserved covariates in Log-normal specification. S1 represents the scenario of Type I group-shared latent heterogeneity due to the influence of discrete group-shared unobserved covariates. S2 refers to the scenario of Type III group-shared latent heterogeneity due to the influence of mixed-type (e.g., both discrete and continuous) group-shared unobserved covariates. S3 represents the scenario of Type II group-shared latent heterogeneity due to the influence of continuous group-shared unobserved covariates. As shown in Figure. 1, the identifiability of subpopulations is gradually reduced from S1 to S3 and the discrete component finally diminishes in S3. The distribution specifications of different structures of group-shared unobserved covariates in Weibull model are similar. To explore various types of group-shared latent heterogeneity, we specify the values of unknown model parameters and use the ground-truth as evaluation benchmarks. Further, we create 12 simulation scenarios in total with different settings of sample size per group *M* and number of groups *n*. We then conduct empirical analysis to investigate all three different structures of group-shared unobserved covariates. We assign less informative priors to all unknown parameters and implement the proposed Bayesian estimation algorithm to estimate model parameters.

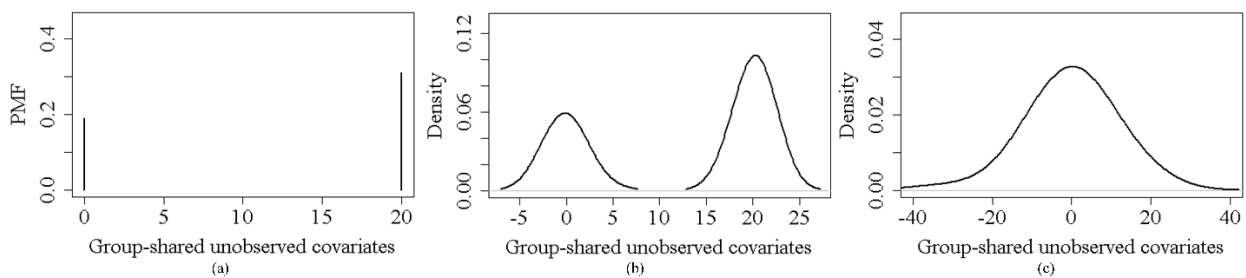

Figure 1: Different types of group-shared unobserved covariates, (a) S1, (b) S2, (c) S3

Based on the estimated models, we use Kaplan-Meier (K-M) survival curves to evaluate and compare the predictive distribution accuracy of proposed approach and alternative modeling approaches. We consider different simulation scenarios and explore both Log-normal and Weibull model specifications in the comparison study. We use the K-M curve calculated from the ground-truth data as benchmark. To investigate the importance of considering group-shared latent heterogeneity, we compare the proposed model with the model which neglects group-shared latent heterogeneity [15]. We train the models and compare the prediction performance of the reliability function evaluated at new values of observed covariates. As illustrated in Figure. 2, predicted reliability curves based on proposed approach for both model specifications in all scenarios are closest to the survival curves of the ground-truth data. On the other

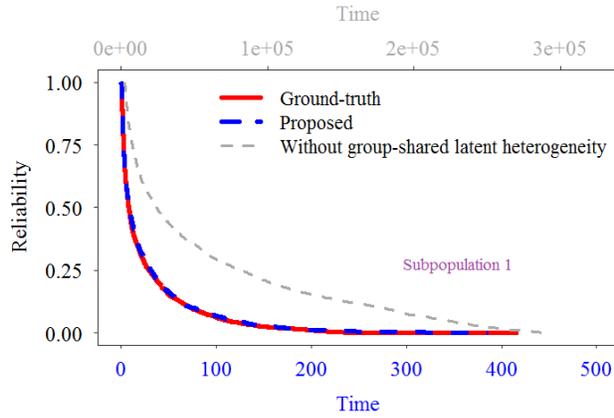
(a) GSLH-D Weibull model in S1

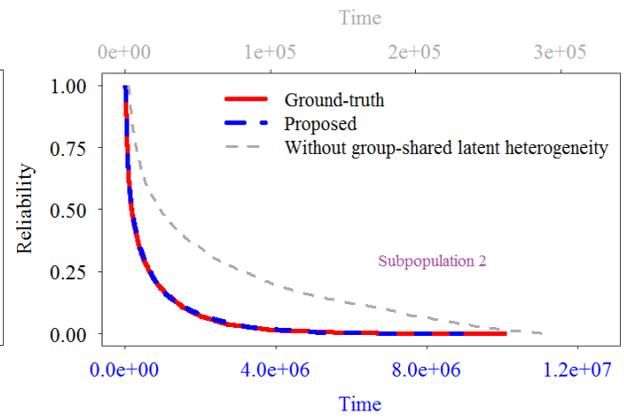
(b) GSLH-D Weibull model in S1

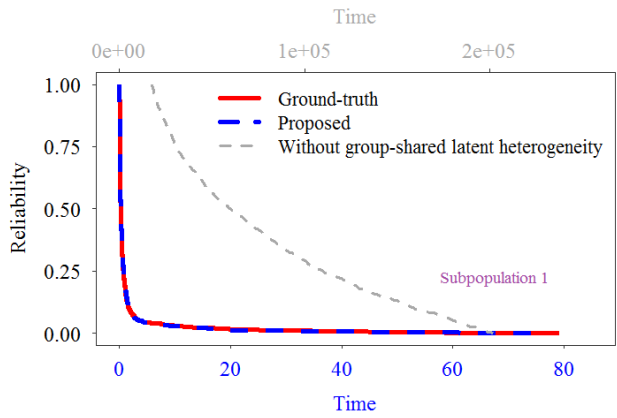
(c) GSLH-M Log-normal model in S2

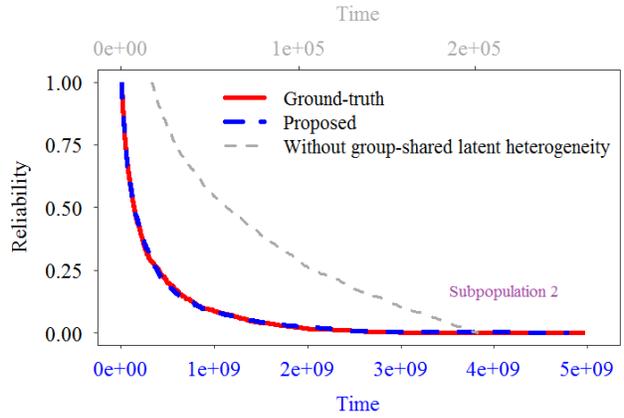
(d) GSLH-M Weibull model in S2

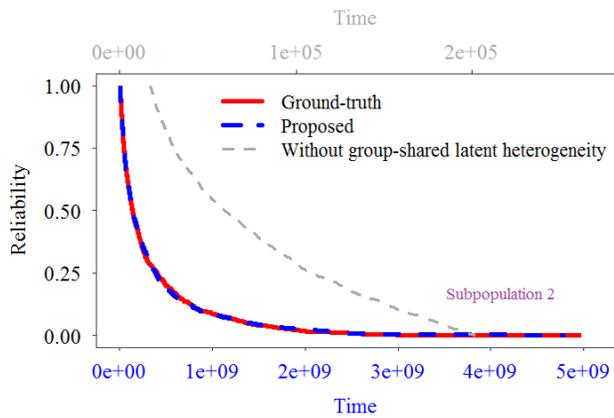
(e) GSLH-C Weibull model in S3

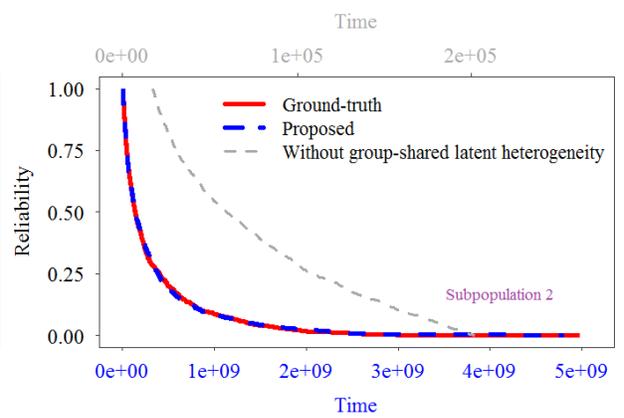
(f) GSLH-C Log-normal model in S3

Figure 2: Predicted reliability function comparison with different specification in different scenarios

side, the model which fails to account for the influence of group-shared unobserved covariates induces either overfitting or underfitting issue for both model specifications in different simulation scenarios, as shown in Figure. 2. Overall, the proposed approach can achieve robust estimates of the group-shared latent heterogeneity and improved prediction performance of reliability function in different scenarios.

Further, we illustrate the estimated group-shared latent heterogeneity in different simulation scenarios. Based on the results of posterior mean and 95% credible interval, the estimated group-shared latent heterogeneities are significantly different from zero, as shown in Figure. 3. This demonstrates that significant influence of group-shared unobserved covariates exists and cannot be neglected. We then use

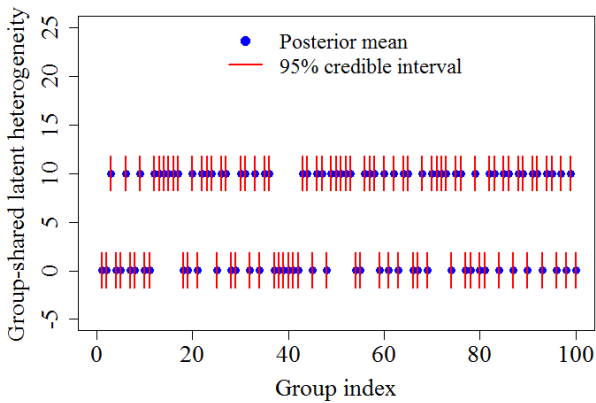

(a) GSLH-D Weibull model in S1

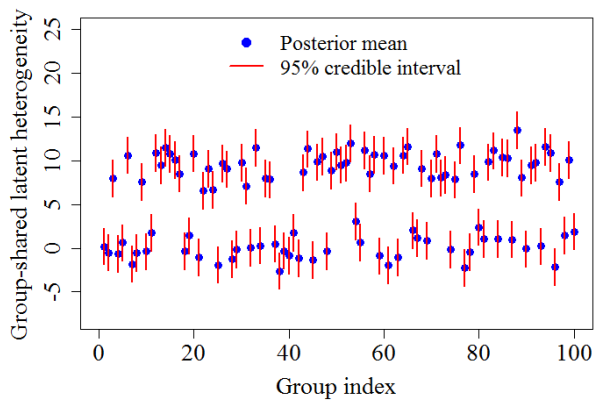

(b) GSLH-M Weibull model in S2

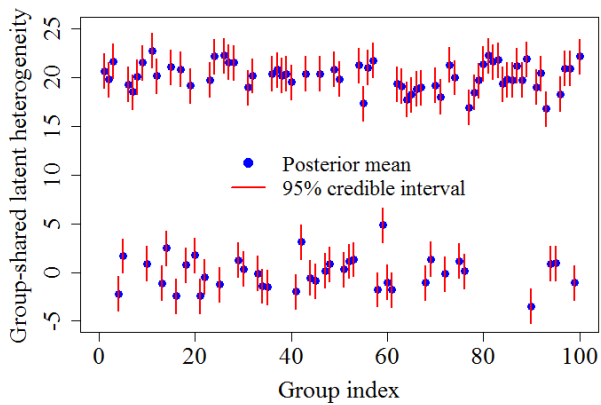

(c) GSLH-M Log-normal model in S2

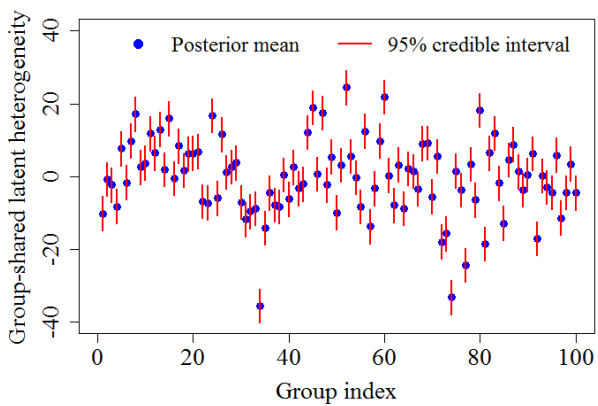

(d) GSLH-C Log-normal model in S3

Figure 3: Posterior mean and 95% credible interval of estimated $W_i$'s with different specifications in different simulation scenarios

GSLH-C model with different simulation settings in scenario S3 as an illustration example to show the density of estimated group-shared latent heterogeneity as well as the ground-truth. As illustrated in Figure. 4, the ground-truth values are fully covered by the derived 95% credible intervals in all cases. When sample size per group $M$ becomes larger, the posterior mode of estimated group-shared latent heterogeneity becomes closer to the true value (illustrated by lines with star symbol in Figure. 4). When number of groups $n$ increases, the precision of estimated group-shared latent heterogeneity is also improved (illustrated by blue lines in Figure. 4).

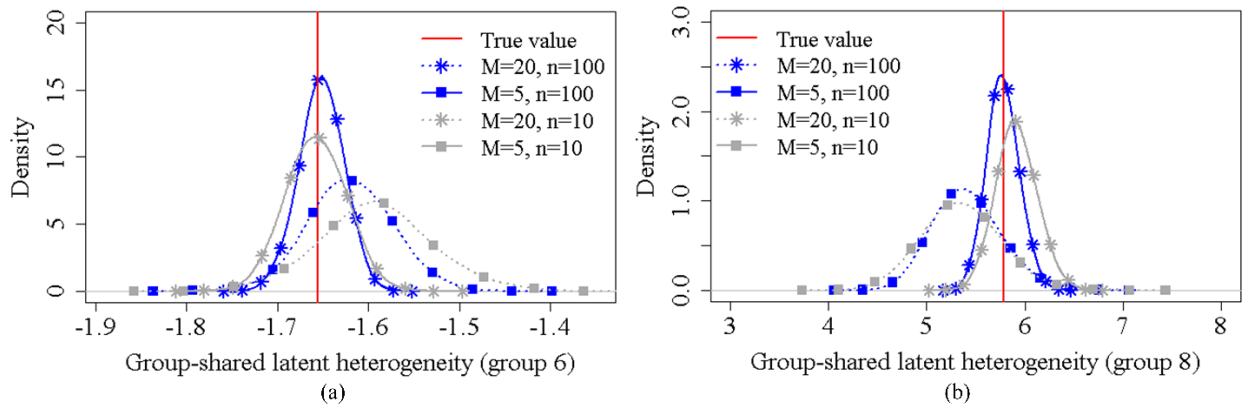

Figure 4: Posterior density plots of group-shared latent heterogeneity based on GSLH-C model in scenario S3: (a) Log-normal, (b) Weibull

Based on the above results, the proposed approach has demonstrated its effectiveness in capturing group-shared latent heterogeneity with improved prediction performance. Further, we present empirical analysis among different simulation scenarios to show the ability of proposed model in identifying the structure of group-shared unobserved covariates. For each simulation scenario, we examine the performance of proposed models with different types of group-shared latent heterogeneity (e.g., GSLH-D, GSLH-C, and GSLH-M model). Deviance Information Criterion (DIC) is used as performance measure to evaluate the goodness-of-fit performance of each model. A smaller DIC value indicates a better goodness-of-fit performance of the evaluated model. In this work, we select the model which completely neglects group-shared latent heterogeneity as baseline model in performance comparison. The empirical comparison results of Log-normal model specification are summarized in Table. 1. The baseline model has worst goodness-of-fit performance in all simulation scenarios with different settings since it fully neglects the group-shared latent heterogeneity. The GSLH-C model among different scenarios achieves better goodness-of-fit performance when different subpopulations become less identifiable and the proportion of continuous component increases. When the continuous component is involved in the group-shared latent heterogeneity (e.g., S2 and S3), both the GSLH-C model and the GSLH-M model achieve improved performance as compared to the GSLH-D model. On the other side, when the group-shared latent heterogeneity is purely discrete, the GSLH-D model achieves the best performance among all models. The adequate model which properly specifies the type of group-shared latent heterogeneity in each scenario can achieve the best goodness-of-fit performance. Based on the performance evaluation results, the proposed approach is able to identify the structure of group-shared unobserved covariates, especially for the discrete type which is not comprehensively addressed in previous studies.

Table 1: DIC comparison results with Log-normal model specification

| Settings | | | GSLH-C | GSLH-D | GSLH-M | Baseline |
|---|---|---|---|---|---|---|
| $S_1$ | $n = 100$ | $M = 20$ | 42986.1 | 42856.6 | 42871.7 | 58751.1 |
| | | $M = 5$ | 10885.2 | 10731.2 | 10766.9 | 14669.1 |
| | $n = 10$ | $M = 20$ | 4984.6 | 4957.2 | 4959.8 | 6527.5 |
| | | $M = 5$ | 1301.2 | 1250.6 | 1252.8 | 1643.1 |
| $S_2$ | $n = 100$ | $M = 20$ | 40624.4 | 51265.9 | 40573.3 | 59196.1 |
| | | $M = 5$ | 10414.1 | 12819.1 | 10244.3 | 14772.9 |
| | $n = 10$ | $M = 20$ | 4867.3 | 5773.1 | 4771.5 | 6614.1 |
| | | $M = 5$ | 1312.4 | 1453.9 | 1213.1 | 1662.8 |
| $S_3$ | $n = 100$ | $M = 20$ | -9996.6 | 6712.2 | -9994.8 | 8475.2 |
| | | $M = 5$ | -2408.3 | 1708.8 | -2402.9 | 2120.9 |
| | $n = 10$ | $M = 20$ | -199.5 | 1307.7 | -199.1 | 1522.4 |
| | | $M = 5$ | -31.1 | 338.7 | -30.2 | 390.9 |

Further, we compare the ability of quantifying the group-shared latent heterogeneity among the GSLH-D, GSLH-C, and GSLH-M models. The posterior mean error, which refers to the average of absolute differences between posterior mean of $W_i$'s and the true values, is calculated to evaluate the estimation

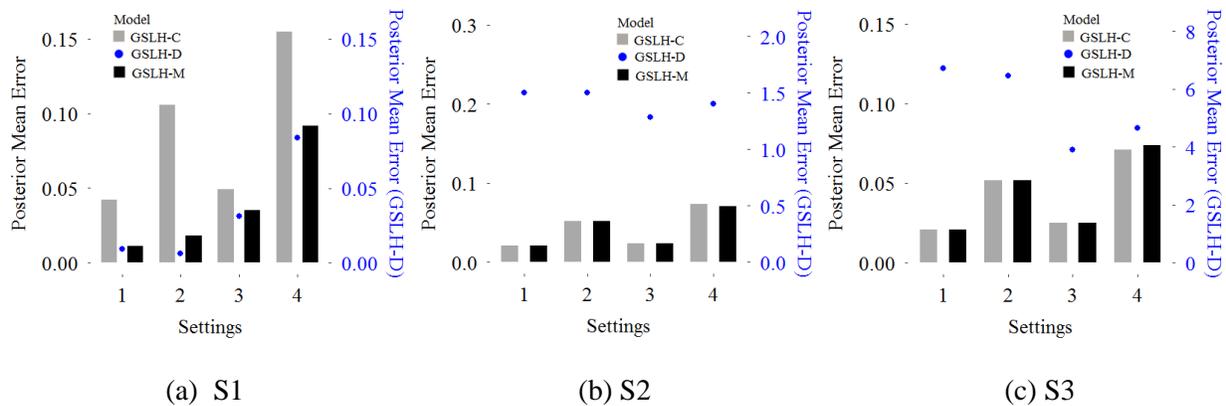

(a) S1  (b) S2  (c) S3

Figure 5: Posterior mean error comparisons of Log-normal specification with different settings (1: M=20, n=100, 2: M=5, n=100, 3: M=20, n=10, 4: M=5, n=10) in different scenarios

performance. The comparison results of posterior mean error based on different models in different simulation scenarios with different settings are illustrated in Figure. 5. Based on the empirical results, the posterior mean errors of the GLSH-D model in S1 are close to the posterior mean errors of the GSLH-M model while the posterior mean errors of the GSLH-C model in S3 are similar as the posterior mean errors of the GSLH-M model. It is because, GSLH-C model and GLSH-D model are special cases of the model with mixed-type group-shared latent heterogeneity. The GLSH-D model achieves smallest posterior mean error in scenario S1 when the group-shared latent heterogeneity is purely discrete type. On the other side, the GSLH-C model results in largest posterior mean error in S1 since it fully mis-specifies the structure of group-shared latent heterogeneity. When the proportion of continuous component increases in the group-shared latent heterogeneity, the posterior mean error of the GLSH-D model becomes significantly larger due to the mis-specification of latent variables. Overall, the GSLH-D model is very sensitive to mis-specification error of group-shared latent heterogeneity while the GSLH-C and GSLH-M models achieve more robust estimates of group-shared latent heterogeneity against the mis-specification error. The results of posterior mean errors based on proposed approach are helpful for identifying the structure of group-shared unobserved covariates and examining whether continuous component is involved in the group-shared latent heterogeneity.

Moreover, the empirical results of Weibull specification are also presented. The model performance in different ground-truth settings of group-shared latent heterogeneity are summarized in Table. 2. Similar

Table 2: DIC comparison results with Weibull model specification

|  | Settings |  | GSLH-C | GSLH-D | GSLH-M | Baseline |
|---|---|---|---|---|---|---|
| $S_1$ | $n=100$ | $M=20$ | 37644.1 | 37498.6 | 37642.2 | 45553.1 |
|  |  | $M=5$ | 9531.1 | 9379.6 | 9522.9 | 11379.1 |
|  | $n=10$ | $M=20$ | 2745.8 | 2734.9 | 2746.9 | 3528.6 |
|  |  | $M=5$ | 689.8 | 678.3 | 690.5 | 877.1 |
| $S_2$ | $n=100$ | $M=20$ | 36826.7 | 39878.5 | 36823.4 | 44836.5 |
|  |  | $M=5$ | 9340.1 | 9966.8 | 9332.7 | 11201.1 |
|  | $n=10$ | $M=20$ | 2470.6 | 2651.7 | 2452.2 | 3207.7 |
|  |  | $M=5$ | 655.2 | 651.8 | 618.1 | 800.5 |
| $S_3$ | $n=100$ | $M=20$ | 12508.1 | 17786.9 | 12509.2 | 19913.3 |
|  |  | $M=5$ | 3247.1 | 4451.1 | 3254.4 | 4961.4 |
|  | $n=10$ | $M=20$ | 564.9 | 879.9 | 565.7 | 1141.9 |
|  |  | $M=5$ | 143.3 | 213.9 | 145.3 | 279.6 |

to the results of Log-normal specification, the GSLH-D model under Weibull specification exhibits better goodness-of-fit performance in scenario S1 where the group-shared latent heterogeneity is discrete type. When the group-shared latent heterogeneity involves continuous component, such as in scenarios S2 and S3, the performance improvements of the GSLH-D model are significantly affected due to the mis-specification of the structure of latent variables. Further, the posterior mean errors of different models under

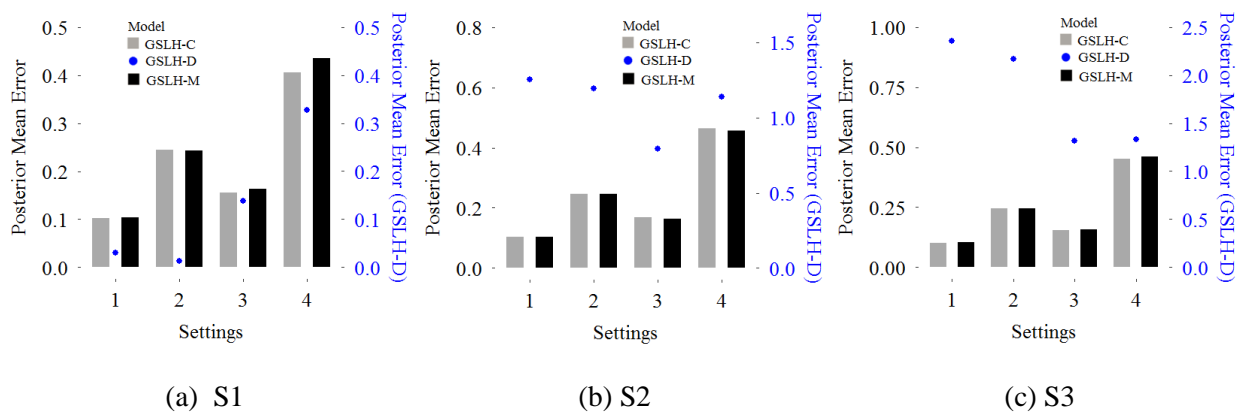

(a) S1  (b) S2  (c) S3

Figure 6: Posterior mean error comparisons of Weibull specification with different settings (1: M=20, n=100, 2: M=5, n=100, 3: M=20, n=10, 4: M=5, n=10) in different scenarios

Weibull specification are illustrated in Figure. 6. Based on the empirical results, the GSLH-D model achieves smallest posterior mean errors in S1 while the posterior mean errors of the GSLH-D model become significantly larger in scenarios S2 and S3 due to the mis-specification of the structures of latent variables. Similar to the results of Log-normal specification, the GSLH-D model is more sensitive to mis-specification error while the GSLH-C and GSLH-M models are more robust to the mis-specification of latent variables in different ground-truth scenarios.

*4.2.Real Case Study*

To further demonstrate the effectiveness and applicability of the proposed modeling approach, we present a real case study on the failure time analysis of test units in tensile tests. In total, 36 tensile tests are performed and the actual lifespans of test units are collected from stress-strain data. To examine the potential group-shared latent heterogeneity, we employ the proposed modeling approach with different types of group-shared latent heterogeneity under Log-normal specification to analyze the lifetime data. We use the model without considering group-shared latent heterogeneity as the baseline model. Without incorporating any covariates information and group-shared latent heterogeneity, the baseline model is simply reduced to Weibull distribution-based model. We then calculate the DIC improvement for each specific model with different types of group-shared latent heterogeneity as compared to the baseline model.

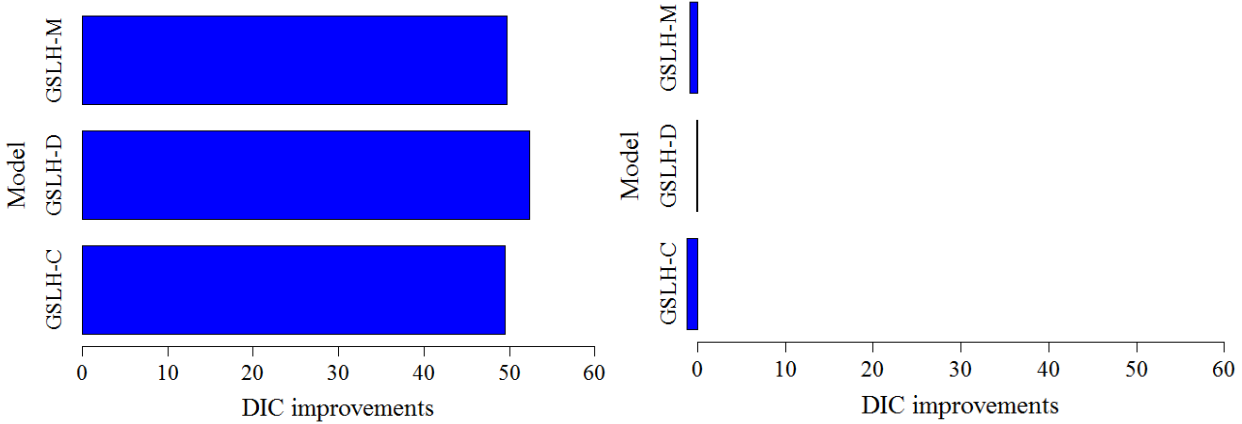

(a) initial study without covariates information　　(b) further exploration with covariates information

Figure 7: Performance comparison of models with different types of group-shared latent heterogeneity

Based on the evaluation results, as shown in Figure. 7 (a), there exists group-shared latent heterogeneity and accounting for such significant influence can improve the goodness-of-fit performance considerably. Further, based on the results of three models with different types of group-shared latent heterogeneity, the GSLH-D model achieves the best performance improvement. This indicates that the group-shared unobserved covariates will be highly likely qualitative. We further extract additional information from the tensile test on each batch based on domain knowledge. We find that different ordinal levels of environmental conditions, such as low and high tensile force as well as low and high temperature, are applied to different batches of test units. This is consistent with the premise that the group-shared unobserved covariates are qualitative. By incorporating such discrete group-shared unobserved covariates as observed covariates, the performance improvement of proposed approach becomes quite small, as shown in Figure. 7 (b). The baseline model becomes Weibull regression model with qualitative observed covariates. With incorporating different types of group-shared latent heterogeneity, there is limited performance improvement of the proposed approach as compared to the baseline model. This indicates that most of the heterogeneity have been explained by the observed covariates and group-shared latent heterogeneity has limited contributions to the model performance improvement. Thus, the baseline model without considering group-shared latent heterogeneity becomes almost equivalent to the proposed approach which accounts for multi-type group-shared latent heterogeneity. This pilot real study demonstrates the abilities of proposed modeling approach in capturing the influences of group-shared unobserved covariates and identifying the structure of group-shared unobserved covariates. The estimation results based on proposed approach can guide the reliability engineer to incorporate the group-shared covariates with identified structure to improve reliability assessment and further to identify the most appropriate design changes with the quantified group-shared latent heterogeneity in design stage. With the improved modeling accuracy, the proposed approach can also facilitate the cost-effective maintenance decisions in the field operation stage.

## 5. Conclusion

In this work, we propose a generic modeling approach to account for the intra-group dependency in product lifetime analysis and to capture the group-shared latent heterogeneity due to the influences of group-shared unobserved covariates. Specifically, we propose a generic lifetime model with different specifications of AFT models to explore different structures of group-shared unobserved covariates and to capture multi-type group-shared latent heterogeneity. Bayesian estimation technique is further employed to jointly estimate the influence of observed covariates and multi-type group-shared latent heterogeneity. Further, we construct different simulation scenarios in numerical study and conduct comprehensive comparison study to demonstrate the improved estimation performance and prediction accuracy of proposed approach. With constructed simulation scenarios, we also present an empirical study to investigate the impacts of group number and sample size per group on estimating the group-shared latent heterogeneity. Moreover, we present real case study to demonstrate the effectiveness of proposed approach. The proposed approach is flexible to capture the different types of group-shared latent heterogeneity and is able to identify different structures of group-shared unobserved covariates.

# Appendix

## A. Proof of Proposition 1

Recall the proposed generic model, as shown in Equation. 2, can be specified with Type I group-shared latent heterogeneity, i.e., $P(W_i = d_k) = p_k, \forall k = 1, ..., K$. Given $W_i$ is known and takes value $d_k$, the lifetime can be quantified by $\log t_{ij} \mid W_i = d_k = \beta_0 + \boldsymbol{\beta}^T \tilde{\boldsymbol{x}}_{ij} + d_k + \epsilon_{ij}$. We let $y_{ij} = \log t_{ij}$. Further, with different assumptions on the distribution of $\epsilon_{ij}$, we can manifest the conditional density $f_y(y_{ij} \mid W_i = d_k)$.

1) When $\epsilon'_{ij}$ follows normal distribution, i.e., $\epsilon'_{ij} = \sigma \eta_{ij}$ where $\eta_{ij}$ follows standard normal distribution and $\sigma > 0$, $y_{ij} \mid W_i = d_k$ has normal density, i.e., $y_{ij} \mid W_i = d_k \sim N(\mu_k = \beta_0 + \boldsymbol{\beta}^T \tilde{\boldsymbol{x}}_{ij} + d_k, \sigma_k^2 = \sigma^2)$. With transformation technique, we can derive the density for $t_{ij} \mid W_i = d_k$ as

$$f_t(t_{ij} \mid W_i = d_k) = f_y(\log t_{ij} \mid W_i = d_k) \cdot (\log t_{ij})'$$
$$= \frac{1}{\sqrt{2\pi}\sigma} \exp\left[-\frac{(\log t_{ij} - (\beta_0 + \boldsymbol{\beta}^T \tilde{\boldsymbol{x}}_{ij} + d_k))^2}{2\sigma^2}\right] \cdot \frac{1}{t_{ij}} = \frac{1}{\sqrt{2\pi}\sigma t_{ij}} \exp\left[-\frac{(\log t_{ij} - (\beta_0 + \boldsymbol{\beta}^T \tilde{\boldsymbol{x}}_{ij} + d_k))^2}{2\sigma^2}\right]$$

Then, we obtain that $t_{ij} \mid W_i = d_k$ follows Log-normal density, i.e., $t_{ij} \mid W_i = d_k \sim LN(\mu_k = \beta_0 + \boldsymbol{\beta}^T \tilde{\boldsymbol{x}}_{ij} + d_k, \sigma_k^2 = \sigma^2)$. Further, we can derive the density of lifetime as

$$f(t) = \sum_{k=1}^K f_t(t \mid W_i = d_k) P(W_i = d_k)$$
$$= \sum_{k=1}^K p_k LN(\mu_k = \beta_0 + \boldsymbol{\beta}^T \tilde{\boldsymbol{x}}_{ij} + d_k, \sigma_k^2 = \sigma^2)$$

Thus, the lifetime can be quantified by log-normal mixture regression model. From the above procedures, we show that when the following conditions hold for the proposed model (see Equation. 2)
(i) with Type I GSLH-D, i.e., $P(W_i = d_k) = p_k, \forall k = 1, ..., K$, and
(ii) $\epsilon_{ij} = \sigma \eta_{ij}$ where $\eta_{ij} \sim N(0, 1)$,
then the Log-normal mixture model with subpopulation-k-specific mean parameter $\mu_k = \beta_0 + \boldsymbol{\beta}^T \tilde{\boldsymbol{x}}_{ij} + d_k$ and variance parameter $\sigma_k^2 = \sigma^2$ is equivalent to the proposed model.

2) When $\epsilon'_{ij}$ follows extreme value distribution, i.e., $\epsilon'_{ij} = \sigma \eta_{ij}$ where $\eta_{ij}$ follows standard gumbel distribution and $\sigma > 0$, $y_{ij} \mid W_i = d_k$ has gumbel density (minimum) with location parameter $\mu_k = \beta_0 + \boldsymbol{\beta}^T \tilde{\boldsymbol{x}}_{ij} + d_k$ and scale parameter $\gamma_k = \sigma$. With transformation technique, we can derive the density for $t_{ij} \mid W_i = d_k$ as

$$f_t(t_{ij} \mid W_i = d_k) = f_y(\log t_{ij}) \cdot (\log t_{ij})'$$
$$= \frac{1}{\sigma} \exp\left[\frac{\log t_{ij} - (\beta_0 + \boldsymbol{\beta}^T \bar{\boldsymbol{x}}_{ij} + d_k)}{\sigma} - \exp(\frac{\log t_{ij} - (\beta_0 + \boldsymbol{\beta}^T \bar{\boldsymbol{x}}_{ij} + d_k)}{\sigma})\right] \cdot \frac{1}{t_{ij}}$$
$$= \frac{1}{\sigma} \frac{1}{t_{ij}} \exp(\frac{\log t_{ij}}{\sigma}) \cdot \exp(-\frac{\beta_0 + \boldsymbol{\beta}^T \bar{\boldsymbol{x}}_{ij} + d_k}{\sigma}) \cdot \exp\left[-\exp(\frac{\log t_{ij}}{\sigma}) \cdot \exp(-\frac{\beta_0 + \boldsymbol{\beta}^T \bar{\boldsymbol{x}}_{ij} + d_k}{\sigma})\right]$$
$$= \frac{1}{\sigma} \cdot t_{ij}^{\frac{1}{\sigma}-1} \cdot \exp(-\frac{\beta_0 + \boldsymbol{\beta}^T \bar{\boldsymbol{x}}_{ij} + d_k}{\sigma}) \cdot \exp\left[-t_{ij}^{\frac{1}{\sigma}} \cdot \exp(-\frac{\beta_0 + \boldsymbol{\beta}^T \bar{\boldsymbol{x}}_{ij} + d_k}{\sigma})\right]$$

Then, we obtain that $t_{ij} \mid W_i = d_k$ follows Weibull density with rate parameter $\lambda_k = \exp(-\frac{\beta_0 + \boldsymbol{\beta}^T \bar{\boldsymbol{x}}_{ij} + d_k}{\sigma})$ and shape parameter $\rho_k = \frac{1}{\sigma}$, i.e., $t_{ij} \mid W_i = d_k \sim Weib(\lambda_k, \rho_k)$. Further, we derive the density of lifetime as

$$f(t) = \sum_{k=1}^{K} f_t(t \mid W_i = d_k) P(W_i = d_k)$$
$$= \sum_{k=1}^{K} p_k Weib(\lambda_k = \exp(-\frac{\beta_0 + \boldsymbol{\beta}^T \bar{\boldsymbol{x}}_{ij} + d_k}{\sigma}), \rho_k = \frac{1}{\sigma})$$

The lifetime can be quantified by Weibull mixture regression model. From the above procedures, we show that when the following conditions hold for the proposed model (see Equation. 2)
(i) with Type I GSLH-D, i.e., $P(W_i = d_k) = p_k, \forall k = 1, ..., K$, and
(ii) $\epsilon_{ij} = \sigma \eta_{ij}$ where $f_\eta(\eta_{ij}) = e^{x-e^x}$,
then the Weibull mixture model with subpopulation-k-specific shape parameter $\rho_k = \frac{1}{\sigma}$ and rate parameter $\lambda_k = \exp(-\frac{\beta_0 + \boldsymbol{\beta}^T \bar{\boldsymbol{x}}_{ij} + d_k}{\sigma})$ is equivalent to the proposed model.